\documentclass[]{elsart}
\usepackage{graphicx}
\usepackage{amssymb}
\begin{document}
\sloppy
\begin{frontmatter}
\title{Gamma-Hadron Separation Methods for the VERITAS Array of Four
Imaging Atmospheric Cherenkov Telescopes }
\author{H. Krawczynski}
\address{Washington University in St. Louis, Physics Department,
1 Brookings Drive, CB 1105, St. Louis, MO 63130}
\ead{krawcz@wuphys.wustl.edu}
\author{D. A. Carter-Lewis}
\address{Department of Physics and Astronomy, Iowa State University, 
Ames, IA 50011-3160, USA}
\author{C. Duke}
\address{Department of Physics, Grinnell College, 
Grinnell, IA 50112-1690, USA}
\author{J. Holder, G. Maier}
\address{School of Physics and Astronomy, University of Leeds, Leeds, 
LS2 9JT, Yorkshire, England, UK}
\author{S. Le Bohec}
\address{High Energy Astrophysics Institute,
University of Utah, Salt Lake City, UT 84112, USA}
\author{G.\ Sembroski}
\address{Department of Physics, Purdue, University, 
West Lafayette, IN 47907, USA}
\begin{abstract}
Ground-based arrays of imaging atmospheric Cherenkov telescopes have emerged 
as the most sensitive $\gamma$-ray detectors in the energy range of about
100 GeV and above. The strengths of these arrays are a very large
effective collection area on the order of 10$^5$ m$^2$, combined with
excellent single photon angular and energy resolutions.
The sensitivity of such detectors is limited by statistical fluctuations 
in the number of Cosmic Ray initiated air showers that resemble gamma-ray 
air showers in many ways. 
In this paper, we study the performance of simple event 
reconstruction methods when applied to simulated data of the 
Very Energetic Radiation Imaging Telescope Array System (VERITAS)
experiment. We review methods for reconstructing the 
arrival direction and the energy of the primary photons, and examine 
means to improve on their performance. 
For a software threshold energy of 300 GeV (100 GeV), the methods achieve  
point source angular and energy resolutions of 
$\sigma_{63\%}\,=$ $0.1^\circ$ ($0.2^\circ$) 
and $\sigma_{68\%}\,=$ 15\% ($22\%$), respectively. 
The main emphasis of the paper is the discussion of $\gamma$-hadron
separation methods for the VERITAS experiment.
We find that the information from several methods can be
combined based on a likelihood ratio approach and the resulting
algorithm achieves a $\gamma$-hadron suppression with a quality 
factor that is substantially higher than that achieved 
with the standard methods used so far.
\end{abstract}
\end{frontmatter}

\section{Introduction}
\label{Introduction}
The combination of the imaging optics and pixilated camera of the Whipple 10 m Cherenkov telescope
and the ``Hillas'' parameterization \cite{Hill:85} of the air shower images led to the initial 
detection of TeV $\gamma$-rays from the Crab Nebula during 1986 to 1988 \cite{Week:89}. 
The discovery established ground-based TeV $\gamma$-ray observations as 
an exciting field at the intersection of astronomy and particle physics. 
Today, almost two decades later, TeV $\gamma$-ray astronomy made
a major impact on our understanding of the blazar class of Active 
Galactic Nuclei \cite{Trev:03,Feli:04,Kraw:05} and is beginning to have a major 
impact on galactic high energy astronomy \cite{Ahar:sc}.

The hardware of ground-based Cherenkov telescopes has evolved significantly. 
Four large Cherenkov observatories have been built or are under
construction: VERITAS \cite{Week:02}, the High Energy Stereoscopic 
Array (H.E.S.S.) \cite{Hint:04}, the Major Atmospheric Gamma Imaging 
Cherenkov detector (MAGIC) \cite{Bast:04}, and CANGAROO III
\cite{Kawa:01}. All these four experiments use (or will use) 
several telescopes to detect air showers in coincidence. 
In the case of the VERITAS array, at first four, then later seven
telescopes are to be located a distance of 80~m from the central telescope.
The H.E.S.S.\ telescope array is already fully operational and consists 
of four telescopes located at the corners of a square with sides 
of 120~m length. In CANGAROO III, the telescopes are located on the
corners of a parallelepiped with sides of 100~m length.
The MAGIC experiment will use two telescopes at a distance of 80~m.
The ``stereoscopic'' detection of air showers with several telescopes 
under different viewing angles suppresses background events initiated 
by muons passing close to individual telescopes and
simplifies the reconstruction of the air shower parameters 
(arrival direction and energy of the primary particle).
While the Whipple 10 m telescope used a camera of 37 pixels 
(0.5$^\circ$ pixel pitch) two decades ago, the modern counterparts use reflectors of 
10~m to 17~m diameter together with cameras containing 500 to
960 pixels ($\sim$0.15$^\circ$ pixel pitch).
These finely pixilated cameras are read out with fast electronics.
In the case of the VERITAS experiment, the signals are sampled at 
500 MHz, making it possible to acquire the air shower 
signals within a few nsec, minimizing the contamination of the 
fast air shower pulses with night sky background photons.\\[2ex]
Stereoscopic observations of air showers with several Cherenkov telescopes requires
a set of analysis methods to reconstruct the air shower parameters, and thus
estimate the arrival direction as well as the type and energy of the primary particle. 
Several different approaches have been used, and we discuss here
simple ``geometrical methods'' \cite{Ahar:97,Kren:98,Hofm:99} 
and so-called ``template fitting methods''  \cite{LeBo:98,Ulri:98}.  
The interested reader might consult \cite{LeBo:05} for a description of
geometric methods specifically aimed at low-energy events.

In the first approach, images are cleaned to suppress signals from night sky 
background photons and characterized with Hillas second moment parameters. 
A Cherenkov image of an air shower looks like a 2-D Gaussian distribution, or
like a ``filled ellipse''; the second moment analysis 
gives the orientation of the major axis of the ellipse and the {\it ``width''} 
and {\it ``length''} parameters. The latter two parameters 
correspond to the root mean square (RMS) of the signal amplitudes perpendicular to the 
major and minor axis of the image, respectively.
Stereoscopic observation of an air shower with several telescopes under significantly 
different viewing angles makes it possible to combine the information from the air shower images 
taken with all the telescopes to reconstruct the orientation of the air shower axis and its location 
relative to the telescopes with a simple geometric method.
Once the air shower axis has been determined, information about the type of the primary particle
(photon or hadron) and its energy can be inferred based on the Hillas parameters.

In the second approach, a dataset of simulated photon-initiated air showers is used to 
derive the expected distribution of signal amplitudes in all telescopes of the experiment
as function of a set of air shower parameters (e.g.\ orientation and location of the 
image, the energy of the primary particle, and the height of the shower maximum).
An air shower event is then analyzed by determining for which parameter combination the
expected signal amplitude distributions best agree with the observed ones.
In other words, the air shower parameters are determined by finding the template image that 
resembles the observed image as closely as possible and minimizes a goodness-of-fit-measure like 
the $\chi^2$-value or a log-likelihood parameter.
The best fit gives a direct estimate of the air shower parameters, and the goodness-of-fit-measure
of the best fit can be used to infer information about the likelihood that the event 
was initiated by a photon rather than a hadron.\\[2ex]
In this paper, we follow the first approach and evaluate its performance when applied to
simulated VERITAS data. 
Direct comparison of the geometrical and template-fitting approaches with the HEGRA
experiment \cite{Daum:97} showed that the first approach can achieve an angular resolution, energy
resolution and $\gamma$-hadron separation capability similar to the 
second \cite{Hofm:99,Ulri:98}\footnote{The CAT collaboration applied a template-fitting method to 
the data of the CAT telescope and reported that the method performed markedly better 
than a simple geometrical method \cite{LeBo:98}.
The difference of the HEGRA and CAT experiences may have several reasons.
It may be that the geometrical method performs relatively better on 
stereoscopic data (HEGRA) than on single telescope data (CAT).
The cameras of the HEGRA and CAT experiments had pixels with angular 
diameters of 0.25$^\circ$ and 0.12$^\circ$, respectively.
The template-fitting method may show its full power only 
when used with a finely pixilated camera.}.
The geometrical approach has the important advantage that it minimizes 
the dependence on the input from Monte Carlo simulations and that it 
is computationally less demanding.
The reconstruction of the air shower axis does not require any Monte Carlo input at all.
The $\gamma$-hadron separation method and the energy reconstruction method 
require minimum simulation input, and experimental air shower data can be 
used to test that the simulations correctly describe the 
relevant air shower properties. 
For example, experimental $\gamma$-ray air shower data
can be used to verify that the simulations correctly predict the lateral 
Cherenkov light distribution \cite{Ahar:99a} as well as the width of the 
images as function of signal amplitude and shower axis distance \cite{Ahar:99b}.
Thus, while the template-fitting approach minimizes the statistical errors on 
the reconstructed air shower parameters, the geometrical method minimizes the 
systematic errors on the analysis results. We think that the robustness 
of the geometrical approach will guarantee its continued use.

\begin{figure}[bh]
\begin{center}
\includegraphics[width=5.7cm,angle=-90]{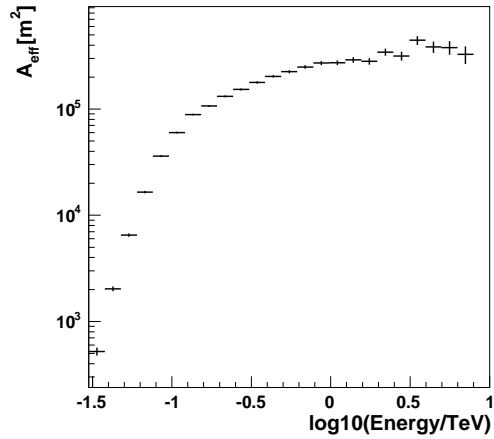}
\end{center}
\caption{\label{aeff} \normalsize Effective area for detecting cosmic $\gamma$-rays as function
of primary energy. Above the threshold region ($>$100~GeV), the effective area increases only
slowly with energy.}
\end{figure}

\begin{figure}[bh]
\begin{center}
\includegraphics[width=5.7cm,angle=-90]{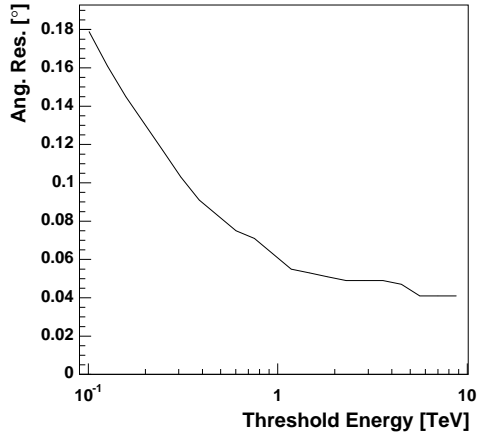}
\end{center}
\caption{\label{angE} \normalsize
\normalsize Angular resolution (63\% value) as function of threshold energy.
The reconstructed energy was used to impose the threshold energy cut.
Only events with reconstructed core locations within 150 m from the 
central telescope were used here. The angular resolution substantially 
improves above the hardware energy threshold.
}
\end{figure}

To limit the scope of the paper we have focused the discussion on the majority 
of the events detected with VERITAS, which means events close to the $\sim$100~GeV 
energy threshold of the experiment. While most of the methods presented in the following 
work well over the entire VERITAS energy range\footnote{Figs.\ \ref{aeff} and 
\ref{angE} show that VERITAS has a good collection efficiency and angular 
resolution at high energies; the ``proton'' curves in Fig.\ \ref{fluxes} 
show that the hadron suppression works well at high energies; finally, 
the ``electron''  curves in Fig.\ \ref{fluxes} show that the cuts accept 
a high fraction of electromagnetic air showers even at high energies.}, 
we have not specifically optimized the methods for giving the best results 
at the high-energy end. The reader is referred to \cite{Ahar:05paris} for a 
discussion of the astrophysics at $>$10 TeV energies, and experimental approaches 
to obtain excellent sensitivities in this energy range.

Another limitation of this paper is that we study only vertically incident
``low zenith angle'' air showers. Literature on the analysis of 
large zenith angle data can be found in \cite{Kren:99,Kono:99,Kosa:04} 
and in the references therein.

The rest of this paper is organized as follows. 
In Sect.\ \ref{SimulatedDataSet} we describe the shower and 
detector simulation codes used, 
and discuss the simulated dataset. Furthermore, we look at 
some raw parameter distributions, like the shower core, direction, and energy 
distributions of the events that triggered the telescope array.
Subsequently, we briefly describe the methods used to reconstruct the air shower 
axis in Sect.\ \ref{ShowerAxis} and the primary energy in Sect.\ \ref{EnergyReconstruction}.
The information about the shower axis reconstruction and the energy 
estimator will be used by the $\gamma$-hadron separation methods.
We address the suppression of Cosmic-Ray initiated air showers in 
Sect.\, \ref{GammaHadronSeparation}. 
The HEGRA and H.E.S.S.\ experiments use a ``scaled width parameter'' 
described further below to separate photons from Cosmic Rays. 
Here, we evaluate several $\gamma$-hadron separation parameters.
The general ideas behind all the $\gamma$-hadron separation methods that 
we will explore were discussed by Hillas and co-workers (see, e.g.\ \cite{Hill:96}) based on very detailed studies of the physical properties of photon and 
Cosmic-Ray initiated air showers. 
Here we study specific implementations of the methods, and evaluate 
their performance when applied to data from the VERITAS experiment. 
A somewhat surprising result from our study is that the information
of the various $\gamma$-hadron separation methods is largely uncorrelated 
and combining the information from several parameters, 
one obtains a much more powerful hadron suppression. 
We conclude with a summary and a discussion in Sect.\ \ref{Discussion}.
\section{Simulation Details}
\label{SimulatedDataSet}
We used the Grinnell-ISU-Utah (GrISUU) air shower and detector simulation 
package\footnote{ http://www.physics.utah.edu/gammaray/GrISU} that combines
the KASCADE air shower simulation code \cite{Kert:94} with the calculation  
of the Cherenkov light emitted by the air shower and the simulation of 
the detector response. We generated 480,000 vertically incident 
$\gamma$-ray-initiated air showers over 
the energy range from 30 GeV to 10 TeV, distributed in energy according to 
a differential power law spectrum $dN/dE\propto E^{-\Gamma}$ with 
photon index $\Gamma\,=$ 2.5. The $\gamma$-rays were assumed to
originate from a point source located at the center of the field of view 
of the telescopes and were simulated over a circular area of 350~m radius.
We simulated 1,930,000 proton-initiated air showers over the energy range from 
100 GeV to 20 TeV with a power law index of 2.7 and with arrival directions 
uniformly distributed over a 4$^\circ$ radius circular area centered on the 
field of view of the telescopes.
The proton-initiated air showers were generated over the same area as the
$\gamma$-ray-initiated ones. All simulations assumed that the 
experiment is located at an altitude of 1.8~km above sea level.
The GrISUU code traces the incoming Cherenkov photons through the
mirror and light cone geometry and simulates the response of the
photomultipliers (PMTs) and the digitization of the signals with the 
VERITAS 500 MHz flash analog to digital converters (FADCs).
It simulates single pixel triggers, and the trigger of a telescope. 
We set the pixel trigger threshold to five photoelectrons. A telescope 
triggers if three pixels fire within 10~nsec and satisfy the pattern 
trigger requirement of any three adjacent pixels.
The telescope array triggers if three telescopes trigger within 50 nsec.
The night-sky noise was set to 4.2$\times10^{11}$ photoelectrons 
Hz m$^{-2}$ sr$^{-1}$.

While the condition of three telescopes triggering in coincidence
results in a higher energy threshold than a two-telescope trigger condition, 
it may optimize the sensitivity of the array for the majority of 
sources detectable with VERITAS. For sources with a soft GeV spectrum, 
a two-telescope trigger may give a higher sensitivity.
When comparing the performance benchmarks of different analysis methods 
(e.g. angular and energy resolution, $\gamma$-hadron separation capability), 
one should be careful to specify the trigger condition. 
The methods discussed below can certainly also be applied to data taken with 
a hardware two-telescope coincidence condition, imposing a software 
requirement of a detected image in at least three telescopes. 
The choice between a two-telescope and a three-telescope trigger 
condition depends also on other considerations, as e.g., the dead time of the data 
acquisition system. It may be the best choice to decide on the trigger condition after 
taking some real data with various trigger conditions, rather than to rely entirely 
on simulations. In the following, we will sometimes use events where at least 
three or at least four telescopes triggered. 
All figures shown in this paper assume the three-telescope trigger condition.

The events were analyzed with the ``eventdisplay'' package developed at 
the University of Leeds. 
The PMT charge is determined by fitting a sample trace to the 
PMT traces and numerically integrating the 
best fit function. 
The code uses pedestal events to determine the average 
pedestal variance (RMS of the pixel charge in the absence 
of an air shower related signal). The pedestal variances
are determined by the rate of night sky background photons and the level of
electronic noise. The analysis characterizes PMT charges in units of 
the pedestal variance of each pixel.
The image cleaning process consists of setting all amplitudes to 
zero for which  neither the ``image'' (the signal amplitude 
exceeds four times the pedestal variance) nor the ``border'' (a pixel adjacent to 
an image pixel with a signal amplitude exceeding two times the 
pedestal variance) condition is met.
\\[2ex]
%
%
In the following we present distributions of several air shower parameters. 
Note that all parameter distributions discussed in this section have been
derived from the true parameter values used in the Monte Carlo simulations. 
Distributions of reconstructed parameters will be discussed in the following sections.
If not stated otherwise, we show distributions for all events that triggered the
experiment and that had at least three telescopes with an image (more than two 
pixels surpassing the image threshold).

\begin{figure}[bh]
\begin{center}
\begin{minipage}{6.5cm}
\includegraphics[width=5.7cm,angle=-90]{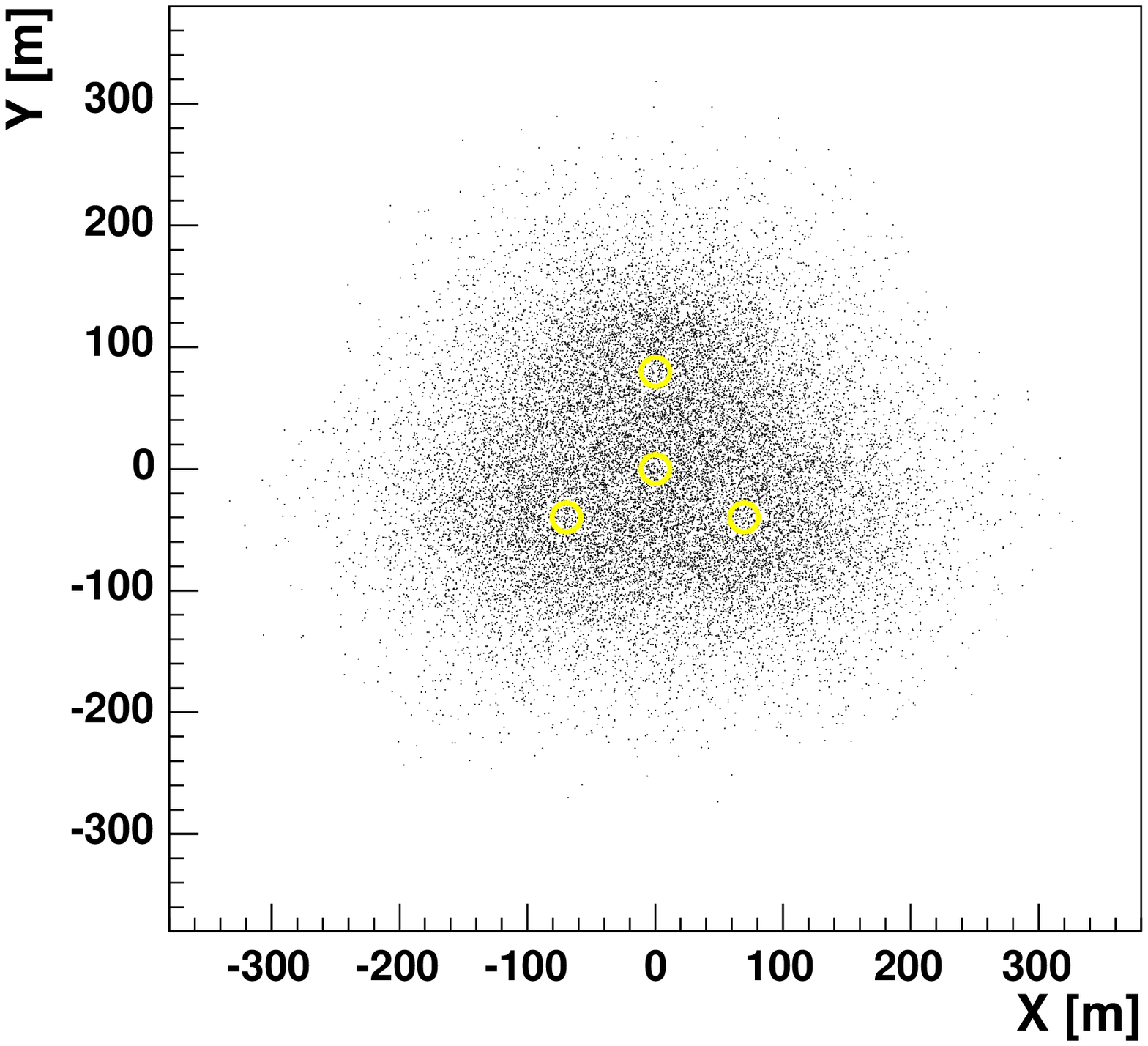}
\end{minipage}
\begin{minipage}{6.5cm}
\includegraphics[width=5.7cm,angle=-90]{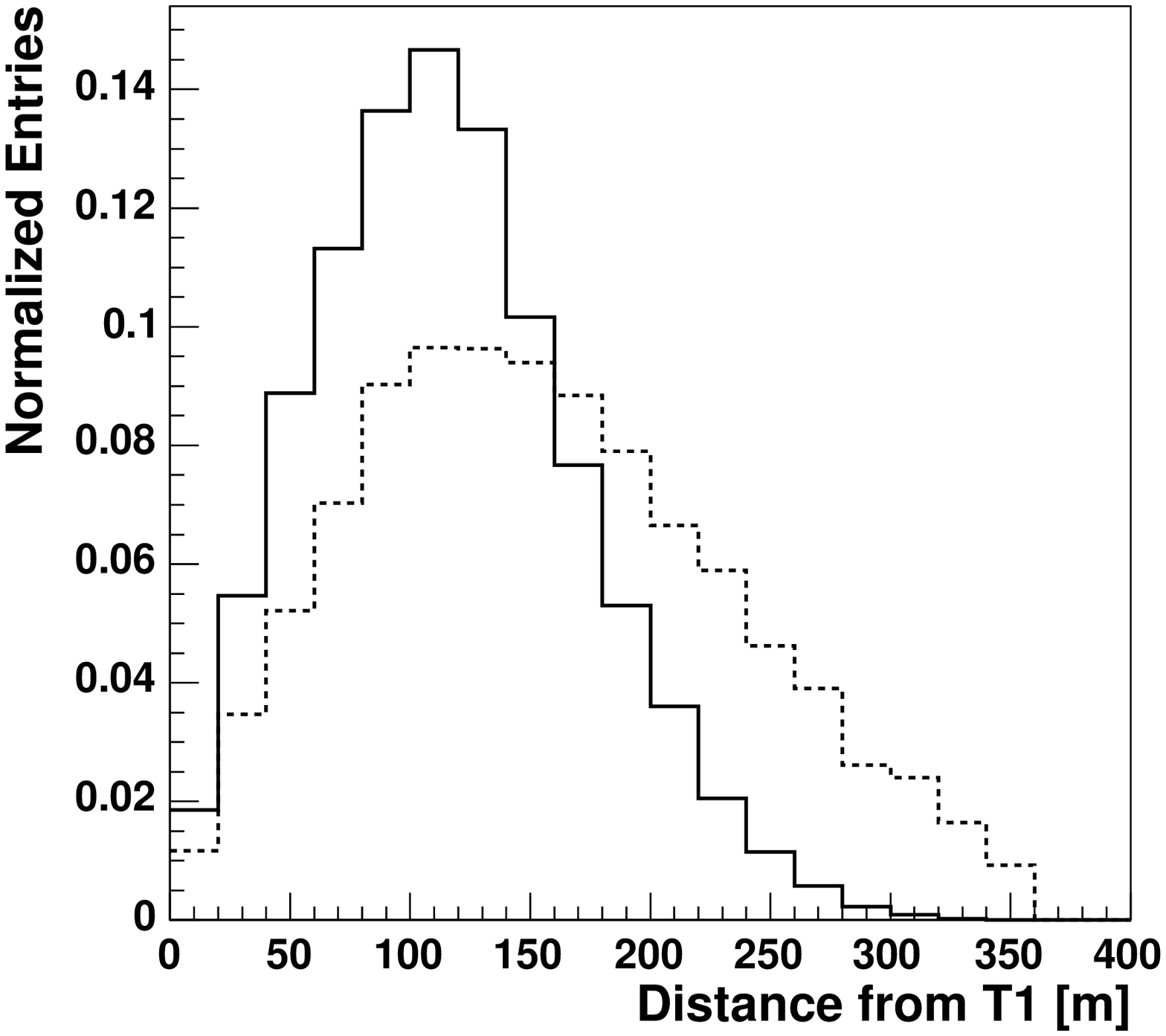}
\end{minipage}
\end{center}
\caption{\label{core1} \normalsize The left panel shows the distribution 
of the core locations of the triggered $\gamma$-ray-initiated
air showers in the plane of the telescopes. 
For reference, the circles give the location of the
4 VERITAS telescopes.
The right panel shows the distribution of the distances 
of the air shower cores from the central telescope T1 
for $\gamma$-ray (solid line) and proton (dashed line) 
initiated air showers that triggered the experiment.}
\end{figure}

\begin{figure}[bh]
\begin{center}
\includegraphics[width=5.7cm,angle=-90]{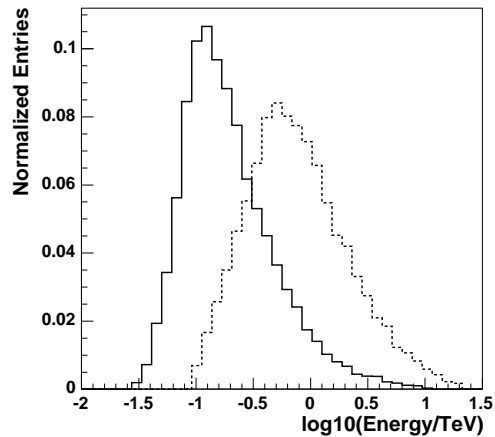}
\end{center}
\caption{\label{energy} \normalsize Distribution of the primary energy of photons (solid line) and protons (dashed line)
that triggered the telescope array. The photon distribution peaks at $\sim$100 GeV and the proton distribution at
$\sim$550 GeV.}
\end{figure}

The left panel of Fig.\ \ref{core1} shows the distribution of air shower cores
of the triggered $\gamma$-rays (the points where the air shower axes 
intersect the telescope plane). Air showers basically 
trigger when their cores fall within the air shower array or less than
$\sim$150~m away from the outer three telescopes.
The right panel of Fig.\ \ref{core1} gives the distribution of the distances 
of air shower cores from the center of the telescope array for both 
triggered $\gamma$-rays and protons. The core distributions 
smoothly approach zero at 350~m from the central telescope, showing that 
we simulated air showers over a sufficiently large area to produce a
realistic air shower dataset.
In Fig.\ \ref{energy} we present the distribution of the energies 
of triggered events for both $\gamma$-rays and protons. The simulations 
cover sufficiently broad energy ranges, as the distributions 
approach zero close to the boundaries of the simulated range.
The differential detection rate per logarithmic energy interval 
peaks at about 100~GeV for photons and 550~GeV for protons.
We see that proton-initiated events have a substantially higher energy 
threshold than $\gamma$-ray-initiated events.
The latter is a well known consequence of (a) hadronic showers 
channeling a fraction of their energy into muons and neutrinos via $\pi^{+,-}$-production, 
and (b) hadronic showers being more irregular than purely electromagnetic showers 
and thus being more unlikely to generate multi-pixel and multi-telescope coincidences.

Fig.\ \ref{aeff} presents the effective detection area for $\gamma$-rays.
It rises proportional to $E^{4.4}$ in the threshold region ($<$100~GeV) and
proportional to $E^{0.27}$ at higher energies  ($>$300~GeV).

\begin{figure}[bh]
\begin{center}
\includegraphics[width=5.7cm,angle=-90]{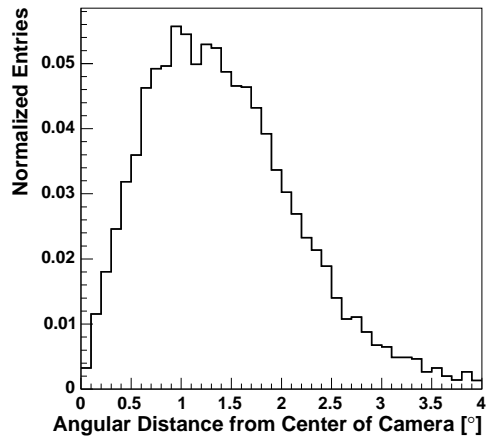}
\end{center}
\caption{\label{fov} \normalsize
 \normalsize Distribution of the angular distance between the arrival directions of all protons that triggered the experiment and the center of the field of view of the cameras. }
\end{figure}

The distance of the arrival directions of triggered proton events from 
the center of the field of view is shown in Fig.\ \ref{fov}. 
At $\sim$4$^\circ$ from the center of the field of view, hardly any
protons trigger the telescopes.
\section{Reconstruction of Shower Axis}
\label{ShowerAxis}
The reconstruction of the air shower direction and core is based on the simple fact
that the major axes of the images are projections of the shower axis.
In the field of view of the telescopes the major axes intersect at 
the point corresponding to the arrival direction of the primary particle.
In the reference frame of the telescopes, all ellipse-like images point 
away from the shower core (the intersection of the shower axis and the telescope plane) 
and the core can again be found at the intersection point of the major axes \cite{Ahar:97,Kren:98}.
If the cosmic $\gamma$-ray source is a point source at a well-known position 
in the sky, the air shower core can be determined with very high statistical 
accuracy. One can use the fact that the lines pointing from the location 
of the source to the centroids of the air shower images point to the air shower core \cite{Hofm:99}.
These lines can be reconstructed with a higher statistical accuracy than 
the major axes of the ellipse images. If not stated otherwise, we will assume 
in the following that the source is a point source of known location.

\begin{figure}[bh]
\begin{center}
\begin{minipage}{6.5cm}
\includegraphics[width=5.7cm,angle=-90]{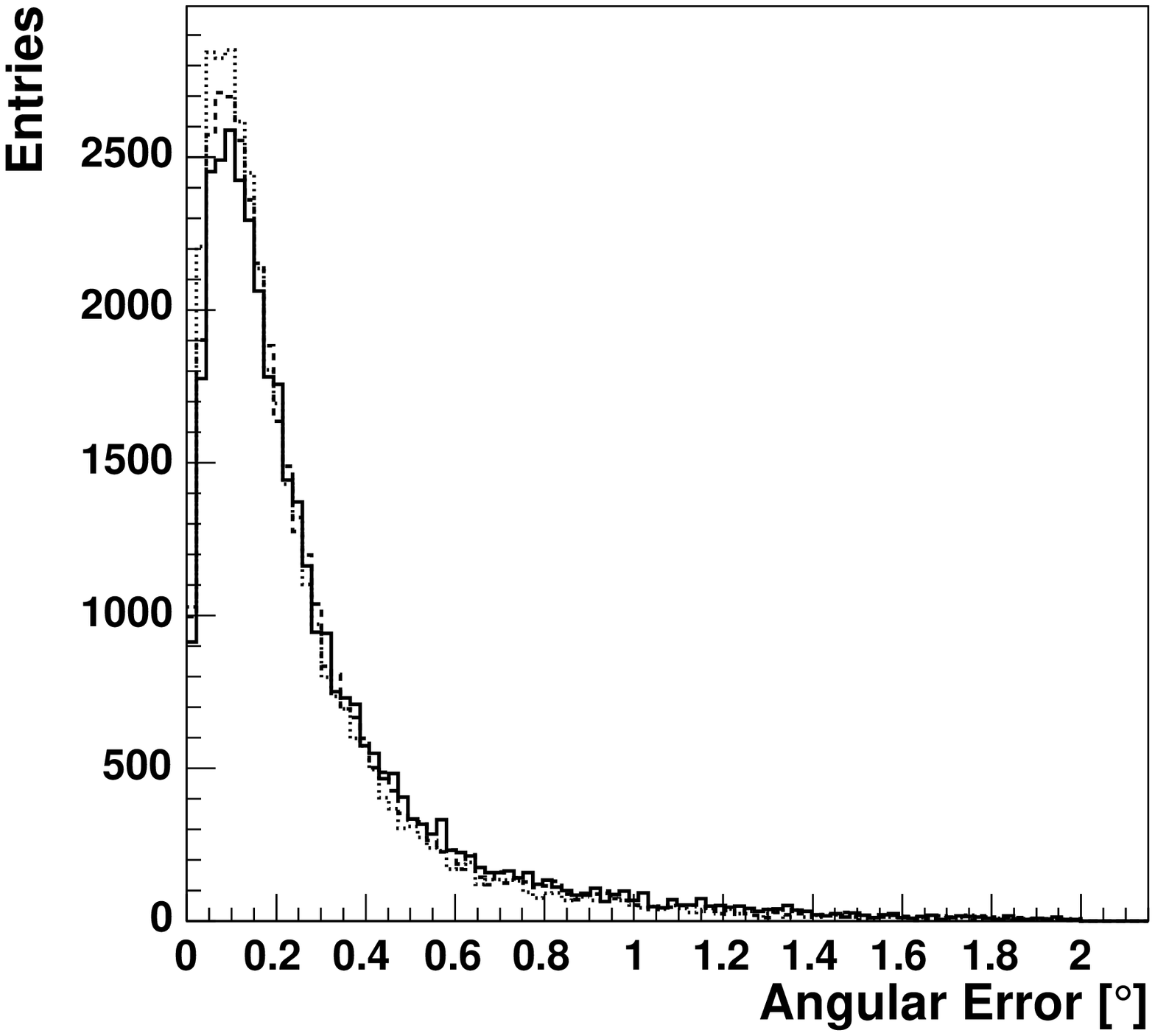}
\end{minipage}
\begin{minipage}{6.5cm}
\includegraphics[width=5.7cm,angle=-90]{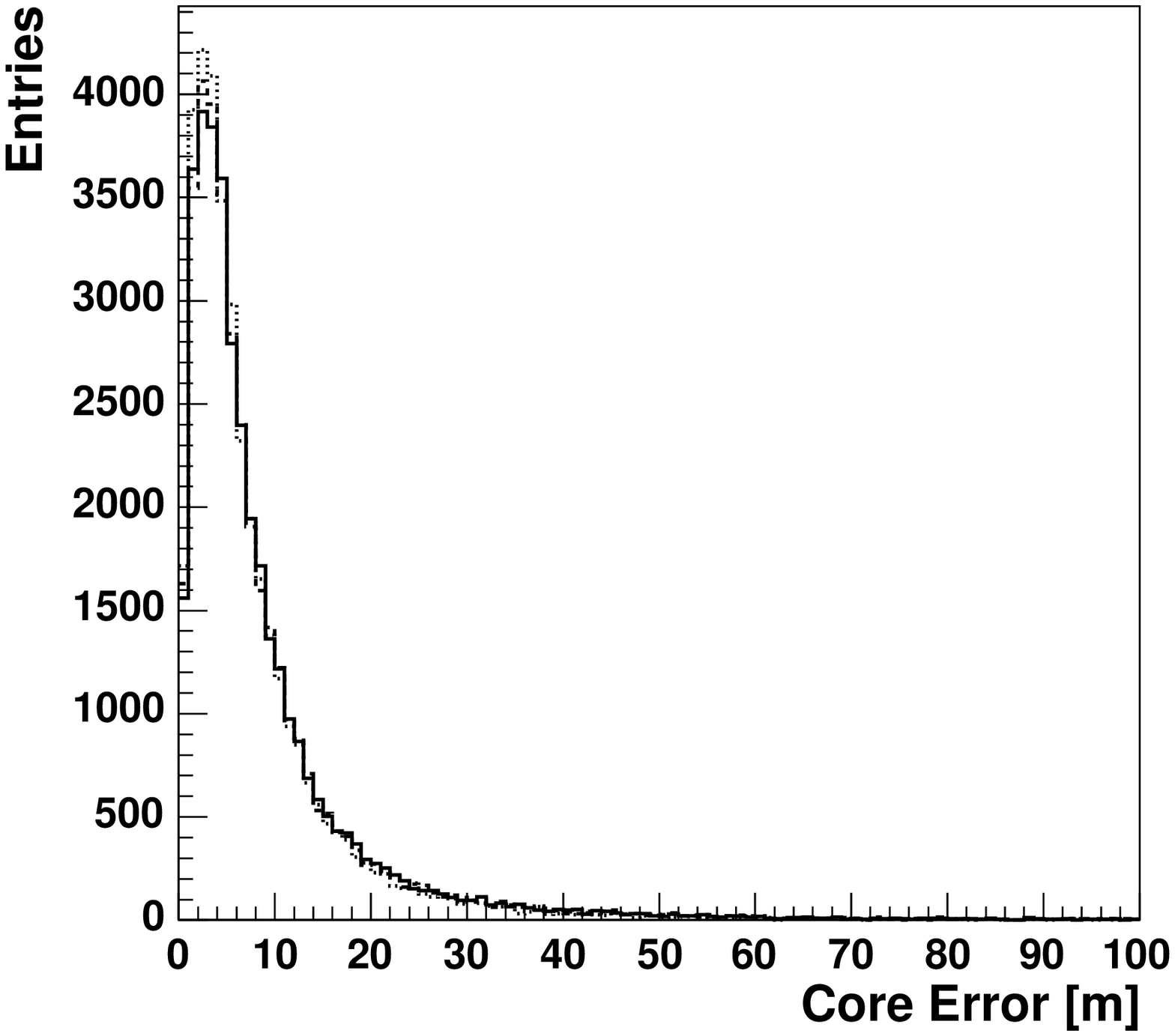}
\end{minipage}
\end{center}
\caption{\label{angres}
The left panel shows the point spread function for all photon-initiated 
events that triggered the telescope array. The different line styles 
show different weighting algorithms. The solid line shows the results 
for constant weights; the dashed line shows weights proportional 
to the {\it size} parameter, and the dotted line
shows a weight that depends on the {\it size}, {\it width} and {\it length} 
parameters. All three weighting schemes give very similar results.
For the best method (dotted line), we get an angular resolution (63\% value) of 0.22$^\circ$.
The right panel shows the distributions of the error in the core 
location for all $\gamma$-ray-initiated events. 
The different line styles correspond to the same 
weighting schemes as describes for the left side.
For the best method (dotted line), we get a core 
resolution (63\% value) of 7.5~m.}
\end{figure}

In practice, finding the best reconstruction algorithm means combining 
the information from all telescopes in an optimal way to minimize the
statistical error of the direction and core estimates.
Here we treat the problem as a $\chi^2$-minimization problem. 
First the direction is determined by finding the point that minimizes 
the weighted squared distances to the major axes.
Subsequently, shower core is reconstructed by minimizing the 
weighted squared distances to the lines that go through the location of the source
and the centroids of the images.
Compared to averaging over intersection points from pairs 
of telescopes, this approach has the advantage that the information about the location 
and orientation of all telescopes enters the estimate at the same time rather than sequentially.
As usual in a $\chi^2$-fit, the weights used for combining the information
of all images should be inversely proportional to the ``squared 
statistical error'' associated with each image. 
We experimented with three different weighting schemes. The simplest ones
use constant weights or weights inversely proportional to the
parameter $size$, the sum of the counts of the corresponding image.
Furthermore, we scrutinized the mean distances between the major axes (or lines 
between the source direction and image centroids) and the 
true direction and core as function of several parameters, 
i.e., $size$, $width$, $length$, $width/length$, and distance 
of the telescope to the shower axis $r$. 
The mean distances indeed scale approximately proportional to $1/size$.
After scaling the errors with $1/size$, we still find a rather strong 
residual dependence on $width$ (not on $width/length$ as one may naively expect
as the $width/length$-values characterize the ellipticities of the images). 
We determined an empirical weighting function that depends on the 
$size$, $width$ and $length$ parameters.
Fig.\ \ref{angres} presents the angular and core
resolutions achieved with three weighting schemes.
While the more sophisticated weighting schemes improve on the angular and core 
resolutions, the improvement is very small.
Weighting according to the third scheme, we obtain an angular and 
core resolution (63 \% value) of 0.22~$^\circ$ and 7.5~m, respectively.
The lack of a substantial improvement may stem from the fact that the images 
in the same shower tend to have the same quality, minimizing the 
effect of the weights. The results agree well with those of an 
earlier study performed for the HEGRA Cherenkov telescope array \cite{Hofm:99}.

The performance improves if high-quality events are selected.
Using only events with a reconstructed shower core within 150~m from
the central telescope and  with a reconstructed 
energy exceeding 300 GeV we obtain angular and core resolutions 
of 0.1$^\circ$ and 4.1~m, respectively. Figure \ref{angE} shows the 
angular resolution for different software energy thresholds. 
The resolution improves with increasing energy threshold and 
above 1 TeV we obtain a resolution of 0.05$^\circ$.
The cut on the core location is critical for obtaining the good results.
Without it, more and more events far away from the telescope array trigger
the array. At the hardware trigger threshold the cut excludes 24\% of the photon initiated
events from the analysis.

For extended sources, we do not know the arrival direction of the photons a 
priori, and the major axes have to be used to determine the shower core location.
Rather independent of the applied weighting scheme, the core resolution deteriorates 
from 7.5~m (4.1~m) for the point source analysis method to 37~m (17~m) for the 
extended source analysis method (the values in brackets have been determined
for a software energy threshold of 300 GeV).
\section{Energy Reconstruction}
\label{EnergyReconstruction}
\begin{figure}[bh]
\begin{center}
\begin{minipage}{6.5cm}
\includegraphics[width=5.7cm,angle=-90]{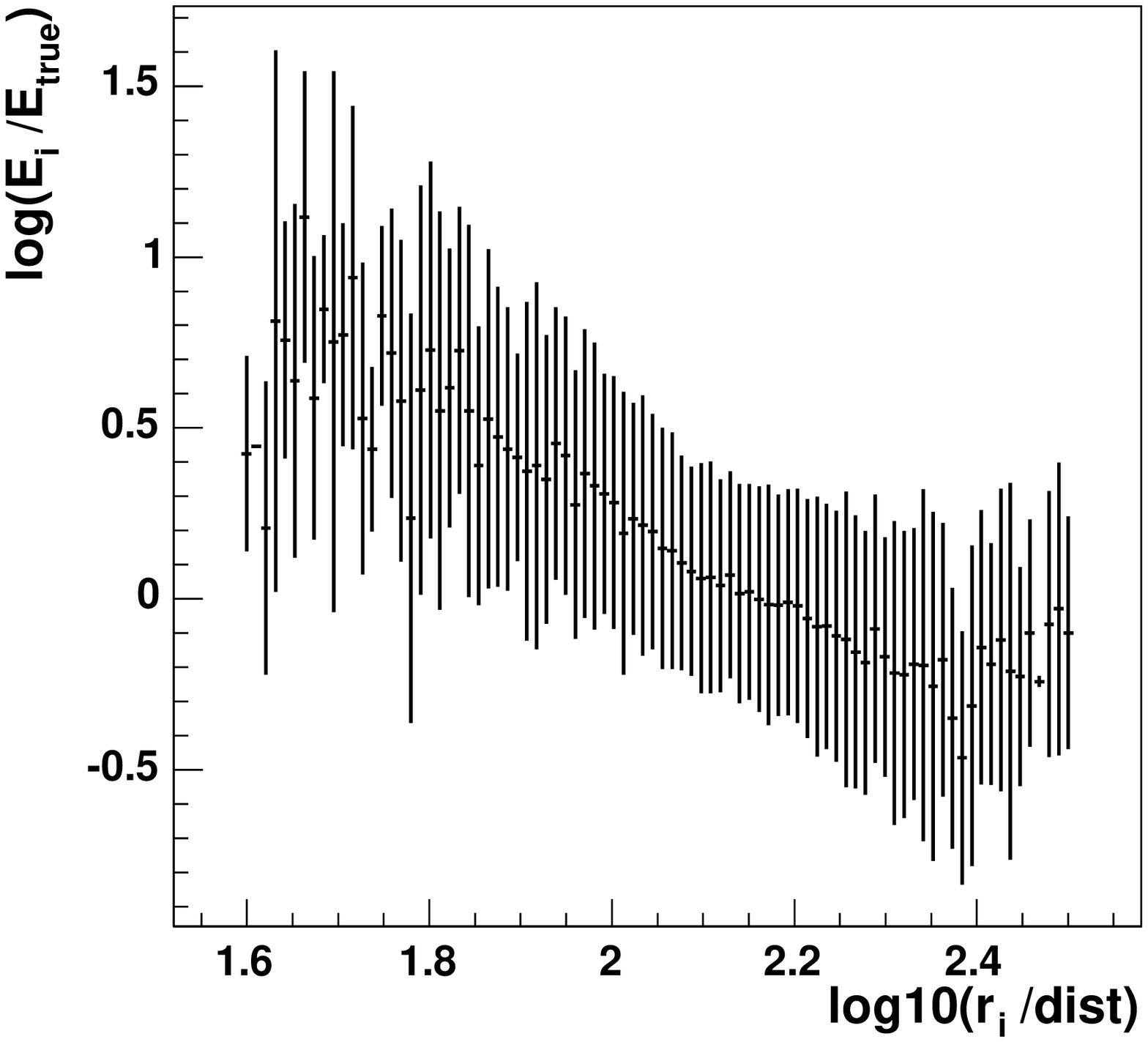}
\end{minipage}
\begin{minipage}{6.5cm}
\includegraphics[width=5.7cm,angle=-90]{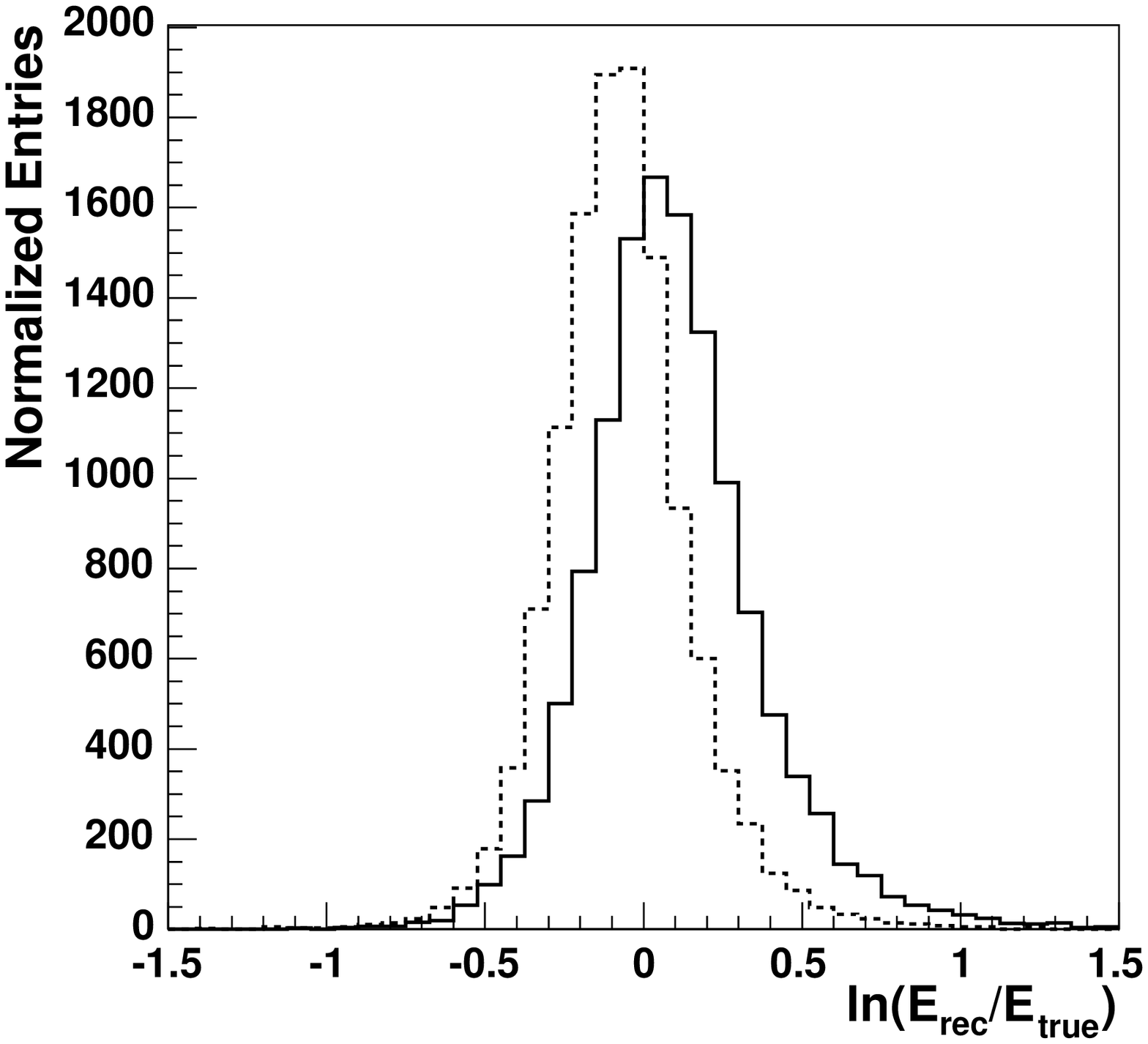}
\end{minipage}
\end{center}
\caption{\label{sh} \normalsize 
The left panel shows the logarithm of the ratio between reconstructed 
and true air shower energy as function of the ratio between the distance 
from a telescope and the {\it dist} parameter measured in that telescope.
The latter ratio depends somewhat on the height of the shower maximum. 
The higher (lower) in the atmosphere the shower develops, 
the larger (smaller) the ratio. One can see that the reconstructed 
energy is overestimated (underestimated) for showers 
developing low (high) in the atmosphere.
The right side compares the energy resolution achieved with 
correction (dashed line) and without  correction (solid line) 
for the height of the shower maximum. 
After (before) correction we get an 
energy resolution of 22\% (28\%). 
The plots include all events for which the reconstructed
energy exceeds 100 GeV and for which the reconstructed shower core less than 200~m away 
from the central telescope.}
\end{figure}

Together, the atmosphere and the Cherenkov telescopes constitute a fully active calorimeter 
with sparse sampling. The energy of the primary particle is roughly proportional 
to the Cherenkov light intensity measured with the telescopes. In the following we discuss 
only the reconstruction of the primary energy for $\gamma$-rays. We implemented a
simple energy reconstruction method (see also \cite{Ahar:99b,Ahar:99c,Ahar:01}).
We determined the median and 90\%-width-values of the logarithm of the {\it size} parameters
as function of the primary $\gamma$-ray energy $E$ and distance $r$ from the shower axis. 
For each telescope with a telescope trigger, an energy estimate is determined by 
inverting the lookup table. 
The energy of the primary particle is determined by averaging the 
energy estimates from all telescope with a telescope trigger.
We weight the estimates from each single telescope proportional to one over the square 
of the statistical uncertainty on the estimate. 
Combining in this way the energy estimates from different telescopes
gives an energy resolution $\sigma_{{\rm ln}{(E_{\rm rec}/E_{\rm true})}}$ 
of 0.28 for the analysis of point sources and 0.4 for the analysis of 
extended sources. We focus in the following on the analysis of point sources.
\\[2ex]
In order to improve on the energy estimate we take into account that the
Cherenkov light intensity depends on the height of the shower maximum\footnote{
A shower reaches the ``shower maximum'' when the number of electrons/positrons 
that emit Cherenkov light reaches a maximum.}.
Given the stereoscopic air shower data, the height of the shower maximum can be
determined. We use the ratio between the distance of a telescope to 
the shower axis $r$ and the angular distance {\it dist} between 
the centroid of the image and the reconstructed shower direction as 
an estimate of the shower maximum. For a constant shower maximum we expect that 
$r$ and {\it dist} are roughly proportional to each other and thus that $r$/{\it dist}\,= {\it const}.
The closer the shower maximum is to the observation level, the further away the image is from the
center of the field of view and the larger is the {\it dist} value for a given 
distance from the shower axis, and the smaller is the $r$/{\it dist} ratio.
The left panel of Fig.~\ref{sh} shows the logarithm of the ratio between the reconstructed energy and 
the true energy as function of $r$/{\it dist}. Indeed, we see that the simple algorithm
severely underestimates (overestimates) the energy for showers with a shower 
maximum high (low) in the atmosphere. We add a correction that depends on the  
$r$/{\it dist} ratio and the {\it size} measured in a telescope.
The dependence on the {\it size}-parameter takes into account that the correction depends somewhat
on the primary energy of the inducing particle.
The right panel of Fig.~\ref{sh} compares the energy resolution achieved with and without the correction.
The correction improves the energy resolution $\sigma_{{\rm ln}{(E_{\rm rec}/E_{\rm true})}}$ of the
air shower array from 0.28 to 0.22.
The performance improves if high-quality events are selected.
Using only events with a reconstructed shower core within 150~m from
the central telescope and selecting events with a reconstructed 
energy exceeding 300~GeV we obtain an energy resolution 
$\sigma_{{\rm ln}{(E_{\rm rec}/E_{\rm true})}}$ of 0.15.
\section{Gamma-Hadron Separation}
\label{GammaHadronSeparation}
One of the main strengths of Cherenkov telescope experiments is the large collection area 
on the order of $10^5$ m$^2$. As a consequence, they achieve unequaled sensitivity 
for $\gamma$-ray observations on short time scales. 
However, for longer integration times their sensitivity is limited by fluctuations 
in the rate of background events and thus increases only with the square root of 
the observation time.
At energies of 100 GeV the air shower background consists of Cosmic Ray hadrons 
and Cosmic Ray electrons. We only discuss the suppression of the first.
Based on today's technology, electron initiated air showers can only be suppressed 
based on their isotropic arrival direction.
In the following we study only proton-initiated air showers. 
Protons are expected to make up about 75\% of the 
Cosmic-Ray initiated background.
The other 25\% are mainly produced by Cosmic Ray He-nuclei. 
We expect that proton-initiated air showers resemble $\gamma$-ray-initiated 
air showers more closely than air showers initiated 
by heavier nuclei.
\subsection*{Normalized Width}
We discuss five methods to distinguish between $\gamma$-ray and hadron-initiated air showers
(see also the excellent discussion in \cite{Hill:96,Fega:96}).
The first approach analyzes the width of the air shower images perpendicular 
to the major axes. Hadron-initiated air showers show significantly ``wider'' images than
$\gamma$-ray-initiated showers owing to the transverse momentum inherent
in hadronic interactions. The transverse momentum originates in the non-negligible 
kinetic energy of the nucleons inside hadronic nuclei and the quarks inside the nucleons.
Similar to Aharonian et al.\ (1997) \cite{Ahar:97}, we use the {\it width} parameters 
measured in all telescopes to derive a $\gamma$-hadron separation parameter.
Based on the Monte Carlo dataset, we derive a lookup table of expected median 
{\it width} values, $w_{\rm m}$, and the 90\%-widths of the distributions, $w_{90}$, 
as functions of the {\it size} of the image and the distance $r$ of the telescope 
from the shower axis.
Here the median and 90\%-width are used rather than the average and the RMS 
to reduce the impact of outliers in the {\it width} distributions.
Here and further below, we have used half the dataset for optimizing the method,
and the other half to measure its performance.

\begin{figure}[bh]
\begin{center}
\begin{minipage}{6.5cm}
\includegraphics[width=5.7cm,angle=-90]{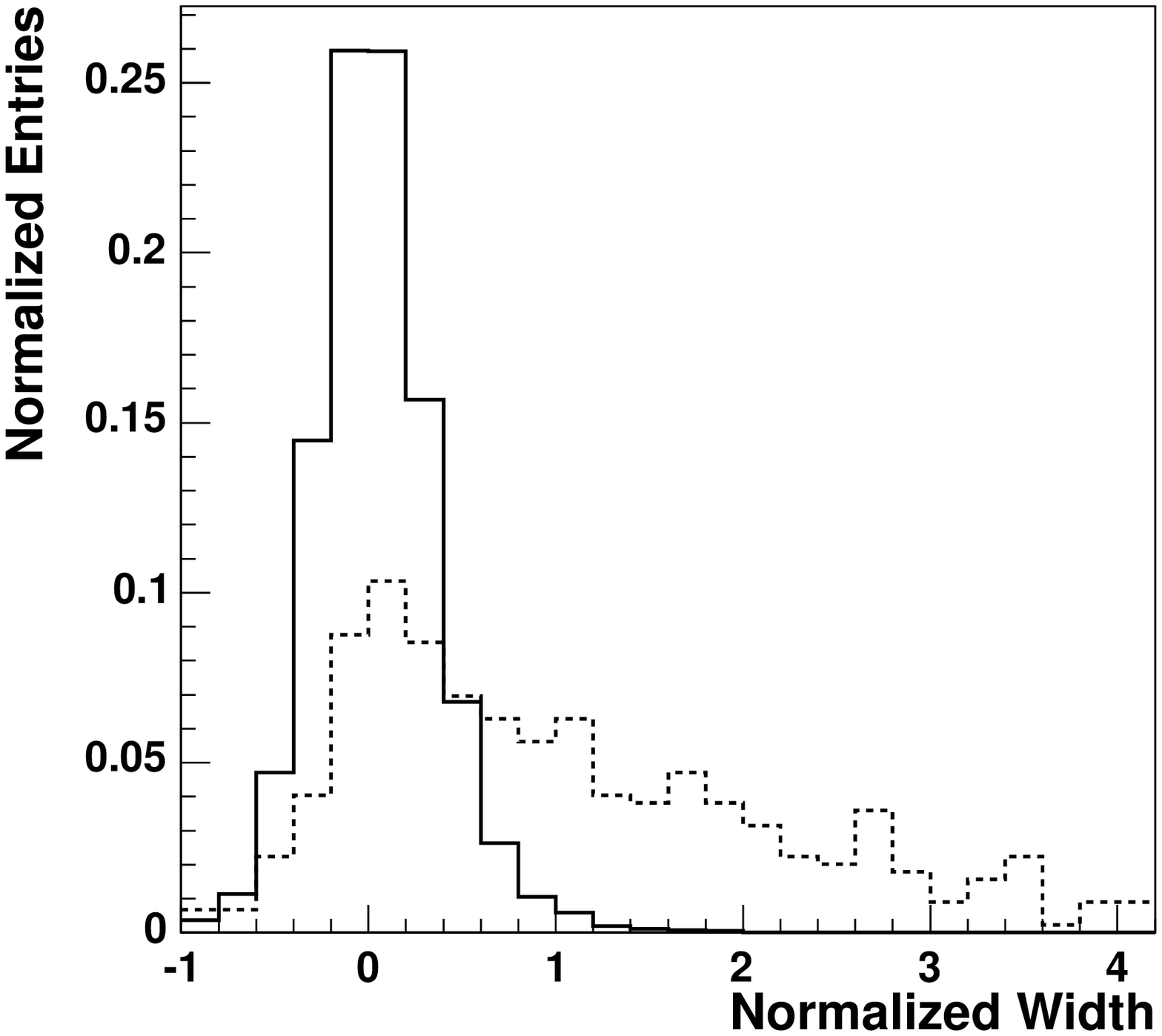}
\end{minipage}
\begin{minipage}{6.5cm}
\includegraphics[width=5.7cm,angle=-90]{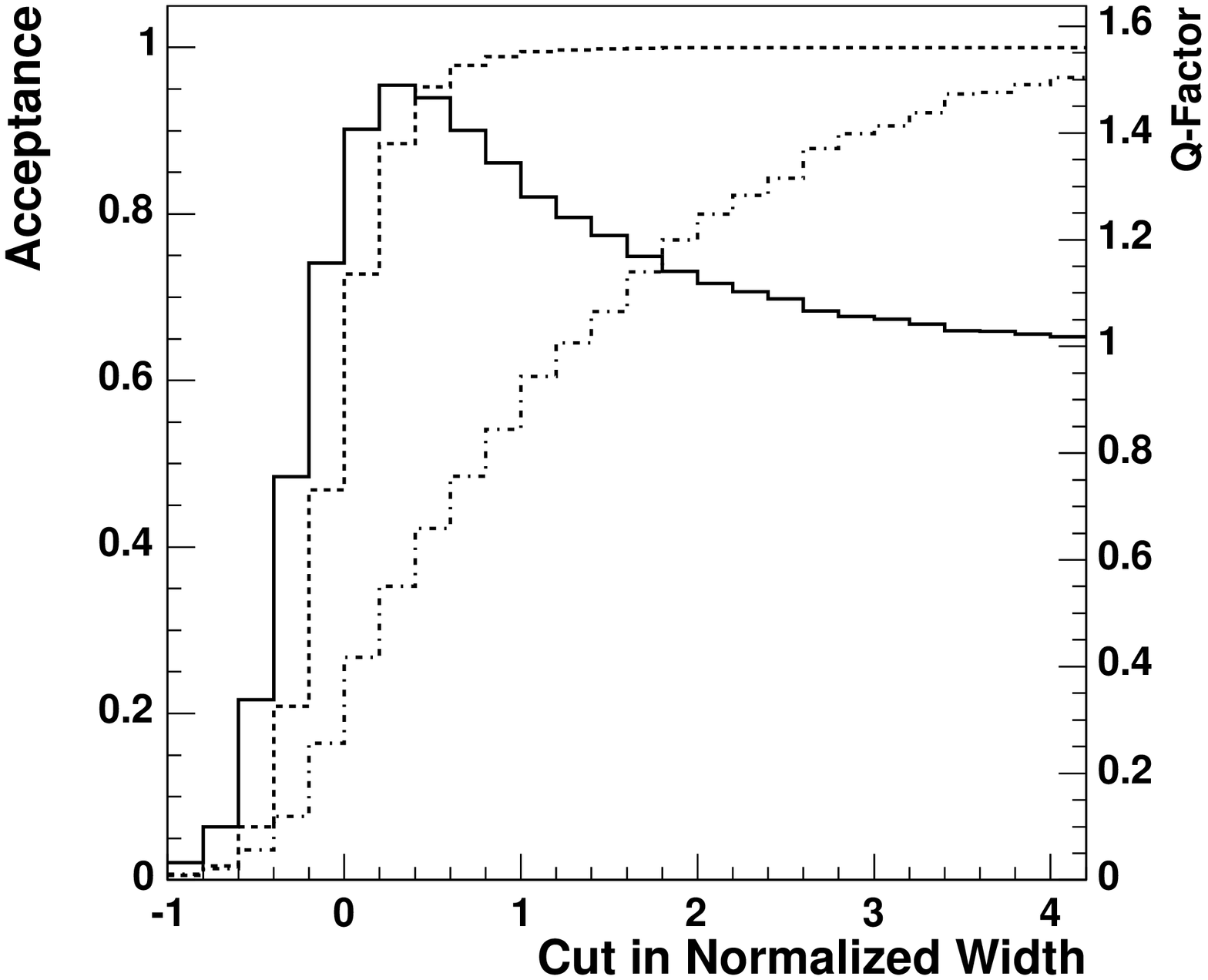}
\end{minipage}
\end{center}
\caption{\label{width} \normalsize 
The left panel shows the distribution of the normalized width parameter for photon (solid line) 
and proton (dashed line) initiated events (three-telescope trigger condition). 
The right panel shows the photon acceptance $\epsilon_\gamma$ (dashed line) 
and proton acceptance $\epsilon_{\rm p}$ (dashed-dotted line) 
as function of a cut on the normalized width parameter $w<w_{\rm cut}$. 
The solid line shows the $Q$-factor. Here it peaks at $Q_{\rm max}\,=$ 1.5. 
}
\end{figure}

Given the $w_{\rm m}(r,size)$ and $w_{\rm 90}(r,size)$-values from the lookup table, 
a ``normalized width'' value is computed by:
\begin{equation}
w\,\,=\,\,\frac{1}{N_{\rm trig}}\times
\left[
\sum_i^{N_{\rm trig}}\,
\frac{width_{\rm i}-w_{\rm m}(r_{\rm i},size_{\rm i})}
{w_{90}(r_{\rm i},size_{\rm i})}
\right]
\end{equation}
where the sum runs over all ${N_{\rm trig}}$ telescopes with a telescope trigger,
{\it width}$_{\rm i}$ and {\it size}$_{\rm i}$ are the {\it width} and {\it size}
values of the image found in the i$^{\rm th}$ telescope with a trigger, 
and $r_{\rm i}$ is the distance of the i$^{\rm th}$ telescope from the shower axis.
The distribution of the normalized width parameter for photons and protons is shown 
in the left panel of Fig.\ \ref{width}. Accepting events with a normalized width 
value below a certain cut value {\it w}$_{\rm cut}$ for the analysis, 
will accept a fraction $\epsilon_\gamma$ of $\gamma$-ray-initiated events and a fraction 
of $\epsilon_{\rm p}$ of proton-initiated events.
The $Q$-factor of a $\gamma$-hadron separation method is defined as
\begin{equation}
Q\,\,=\,\,
\frac{\epsilon_\gamma}{\sqrt{\epsilon_{\rm p}}}
\end{equation}
where $\epsilon_\gamma$ and $\epsilon_{\rm p}$ are the fractions of 
photon and hadron
initiated events surviving the cut.
The $Q$-factor resembles the improvement in signal to noise ratio achieved
with a cut, assuming that the noise is dominated by background 
fluctuations. 
The values $\epsilon_\gamma$, $\epsilon_{\rm p}$ and the $Q$-factor are given 
as function of the cut value {\it width}$_{\rm cut}$ in the right panel of Fig.\ \ref{width}.
We obtain an optimal $Q$-factor of 1.5 for $w_{\rm cut}\,=$ 0.3.
Using the more restrictive condition that four rather than three telescopes triggered, improves
the performance of the normalized width cut to a maximum $Q$-value of 2.38 for 
$w_{\rm cut}\,=$ -0.1.
The performance of the normalized width cut, as well as that of the $\gamma$-hadron
separation methods described in the following paragraphs is summarized 
in Table~\ref{sum}.
\subsection*{Agreement Between Different Telescopes Regarding the Shower Direction and Core}
Krennrich \& Lamb (1995) \cite{Kren:95} and Hillas (1996) \cite{Hill:96} 
suggested to use the deviation of the major axes from the reconstructed 
shower direction and/or core location as means to distinguish between photon and hadron-initiated events. 
The wider and more irregular hadron-initiated showers tend to produce images that do not all 
point to the arrival direction and core location.
We use chi-square values $\chi_{\rm dir}^2$ and $\chi_{\rm core}^2$ 
of the direction and core fits, respectively, to differentiate between 
photons and hadrons. We normalize the chi-square values according
to the number of telescopes participating in the fit.
We find that the optimal cuts in the chi-square values $\chi_{\rm dir}^2$ 
and $\chi_{\rm core}^2$ result in hadron suppressions with 
$Q$-factors of 1.34 and 1.69 (three-telescope trigger condition) and 
1.53 and 2.3 (four-telescope trigger condition), respectively.
The cut in $\chi_{\rm core}^2$ is very powerful, and achieves a comparable
$\gamma$/hadron separation as the cut in the normalized width parameter.
\subsection*{Lateral Cherenkov Light Distribution}
The fourth method takes into account that photon and hadron initiated air showers
exhibit a different lateral Cherenkov light distribution, and that Cherenkov light pool
is more homogeneous for photon initiated air showers than for hadron initiated air showers.
We make use of these differences by comparing the energy estimates derived from the 
information of individual telescopes with the energy derived from the information of all telescopes.
Hereby, all energy estimates assume that the primary particle is a photon.
We define a reduced chi-square value as follows:
\begin{equation}
\chi_{\rm E}^2\,\,=\frac{const}{N_{\rm trig}-1} 
\times 
\sum_{i=1}^{N_{\rm trig}}\,\,\frac{{\rm ln}(E_{\rm i}/E_{\rm all})}{\sigma_{{\rm ln}{(E_{\rm i})}} ^{\,\,2}}
\end{equation}
The sum runs over all telescopes with a trigger, $E_{\rm all}$ is the energy estimate derived from all telescopes,
$E_i$ is the energy estimate derived from the i$^{th}$ telescope, and $\sigma_{{\rm ln}{(E_{\rm i})}}$ is the 
estimated error on ${\rm ln}{(E_{\rm i})}$.
Optimizing the cut-value, we find that the $\chi_{\rm E}^2$-cut gives
maximum $Q$-factors of 1.30 and 1.60 for the three-telescope and 
four-telescope trigger conditions, respectively.
\subsection*{Making Use of the Temporal Information}
Another consequence of the more regular nature of the electromagnetic showers compared to 
hadronic showers, is that their Cherenkov light front is narrower in time.
The VERITAS array of Cherenkov telescopes is equipped with fast FADCs to read out the pixel signals.
As fifth $\gamma$-hadron separation method, we compute for each pixel the time at which the signal rose to 50\% of its
peak, and compute a reduced chi-square value $\chi_{\rm time}^2$ 
as the root mean square (RMS) value of these rise times divided 
by the degrees of freedom. 
For the three-telescope and four-telescope trigger conditions, 
a cut in $\chi_{\rm time}^2$ gives $Q$-factors of 
1.09 and 1.22, respectively.
\subsection*{Combining the Information of Various Methods}
\begin{figure}[bh]
\begin{center}
\begin{minipage}{6.5cm}
\includegraphics[width=5.7cm,angle=-90]{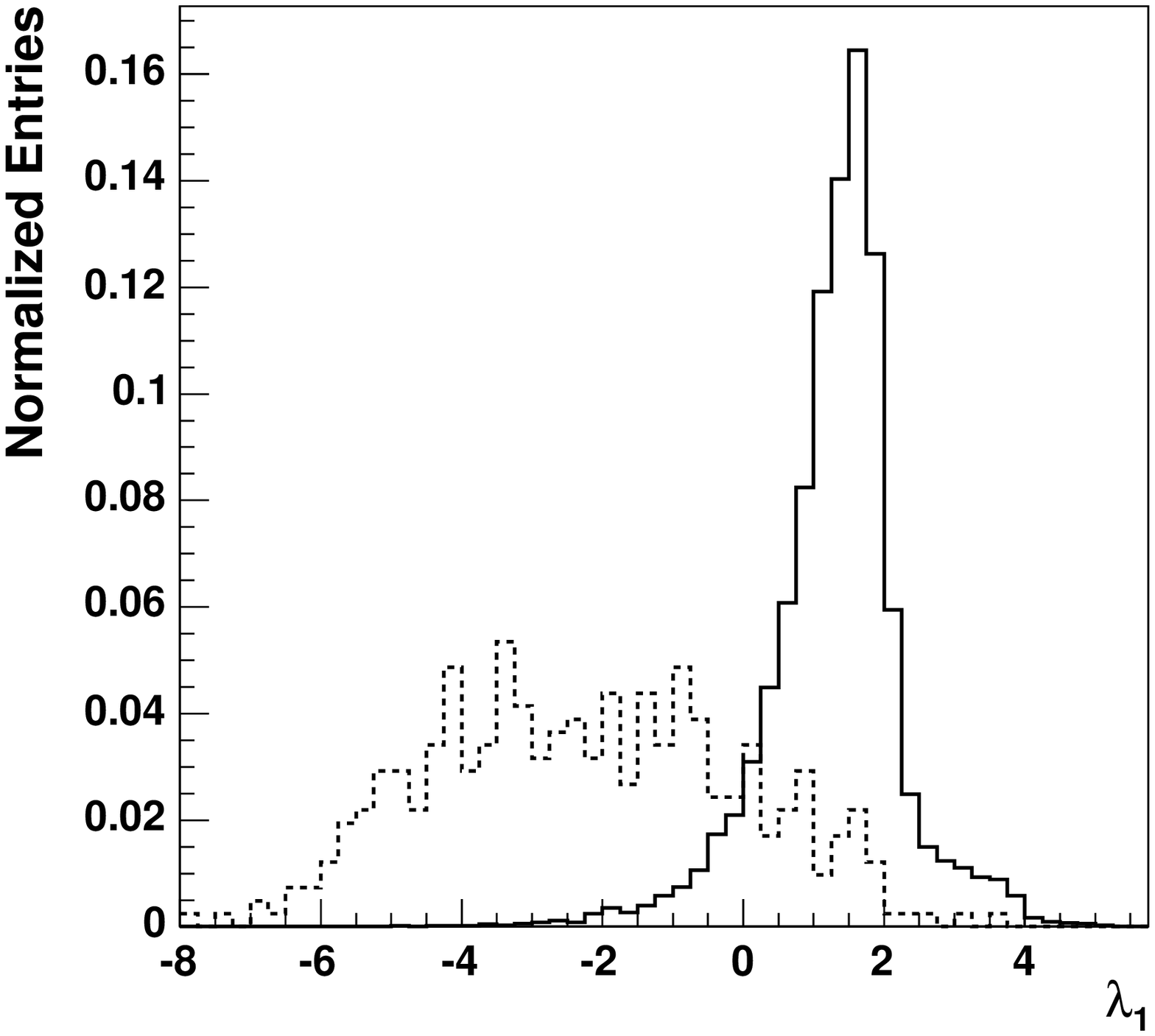}
\end{minipage}
\begin{minipage}{6.5cm}
\includegraphics[width=5.7cm,angle=-90]{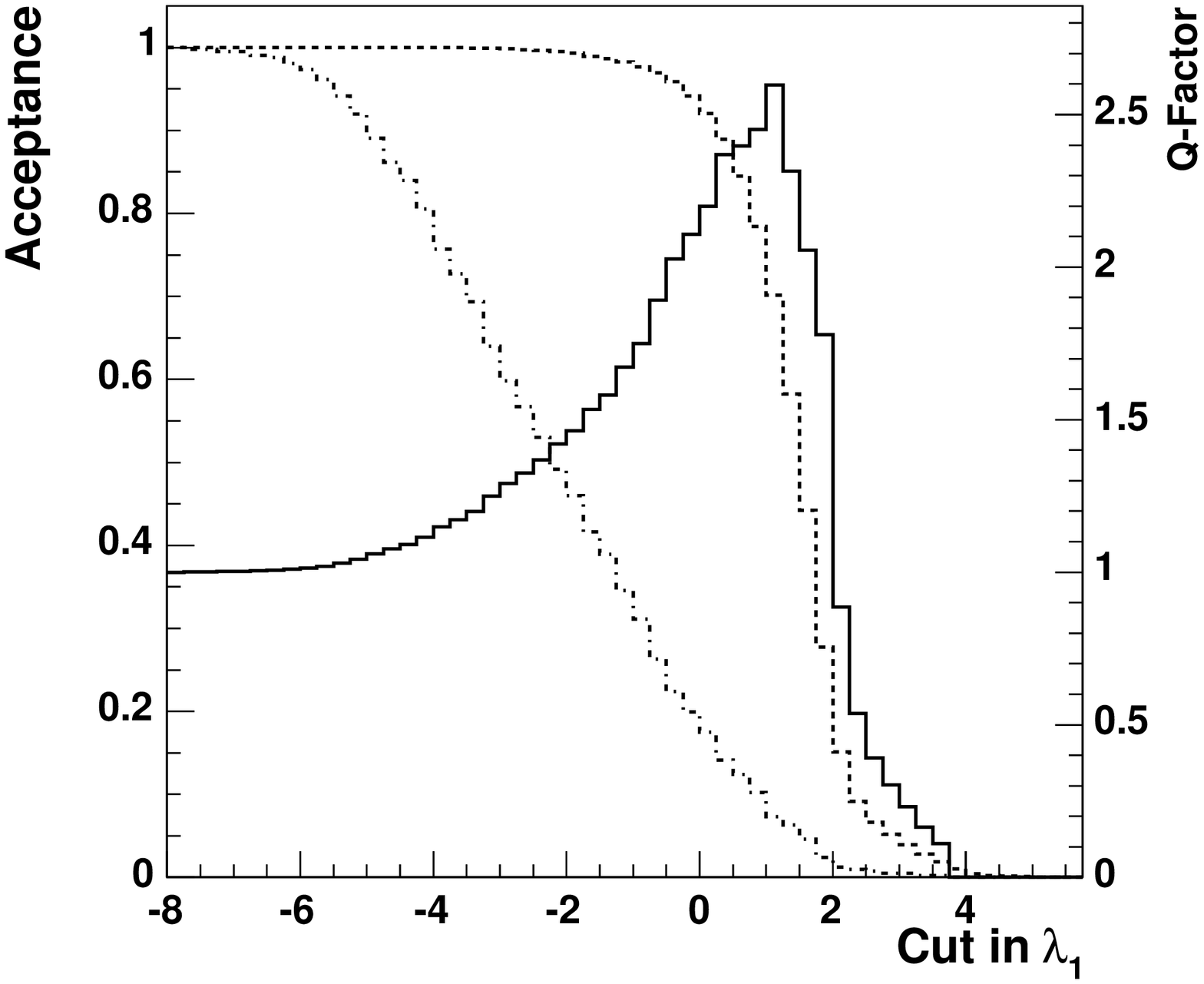}
\end{minipage}
\end{center}
\caption{\label{ml2} \normalsize
Same as Fig.\ \ref{width} for the cut
$\lambda_1(w,\chi_{\rm dir}^{2},\chi_{\rm core}^{2},\chi_{\rm E}^{2})$.
The maximum $Q$-factor is 2.60.
}
\end{figure}

%
%
Scatter plots between two of the five $\gamma$-hadron parameters discussed above show that the parameters
are not strongly correlated with each other. Cutting in several parameters should result in an 
improved performance. We apply the likelihood ratio formalism to combine the information 
from different approaches. We use the $\gamma$-ray and proton distributions of the cut parameters 
as probability density functions (PDFs).
The $\gamma$-hadron separation parameter is then the logarithm of the ratio of the probability
that the event is a photon divided by the probability that it is a proton.
We define the parameter:
\begin{equation}
\label{l1}
\lambda_1\,\,=\,\, \sum_{i=1}^{N}\left[ {\rm ln} (P_{\rm i}^{\gamma})- {\rm ln}(P_{\rm i}^{\rm p})\right]
\end{equation} 
where the sum runs over the $N$ $\gamma$-hadron separation methods to be used, and
$P_{\rm i}^\gamma$ and $P_{\rm i}^{\rm p}$ are the probabilities that photon and 
proton-initiated air showers produce the observed value of the i$^{th}$ parameter, respectively.
While this approach should give the optimal results if the parameters are indeed
completely uncorrelated, we also compute a second parameter which might work better if there
is some correlation between the parameters:
\begin{equation}
\label{l2}
\lambda_2\,\,=\,\, \min_{i=1..N} \left( {\rm ln}(P_{\rm i}^\gamma)- {\rm ln}(P_{\rm i}^{\rm p}) \right)
\end{equation} 
Basically, we identify an event as a proton-initiated event if one of the used
parameters strongly favors the proton over the photon hypothesis.
The performance of several combinations of the separation parameters are given in Table 1.
For the three-telescope coincidence condition, we find that combining the information from
all five $\gamma$-hadron separation methods improves on the best $Q$-factor 1.7 of a 
single method, to a $Q$-factor of about 2.6 (see Fig.\ \ref{ml2}). 
In the case of the four-telescope trigger condition, the parameter $w$ alone 
gives $Q\,=$2.4. Combining the information from $w$ and $\chi^2_{\rm core}$ 
improves the performance to a $Q$-factor of 3.6.
In all cases, there is little difference between using Equ.\ (\ref{l1}) 
or Equ.\ (\ref{l2}) for combining the information.

Please note that we get very comparable results and almost identical 
$\gamma$-hadron 
separation $Q$-factors for the analysis of point sources and extended sources.

Finally, we would like to mention that imposing the four-telescopes 
trigger condition reduces the number of detected $\gamma$-ray and cosmic 
ray events. Taking into account the numbers of detected events, the 
achievable angular resolutions, and the best $Q$-factors, we compute that 
the signal to noise ratio of a point source detection is by a factor 
of 1.9 better for the events taken with the four-telescope trigger 
condition than for all the events
taken with the three-telescope trigger condition. 
\section{Summary and Discussion}
\label{Discussion}
We have used a dataset of simulated photon and proton air showers to
study the performance of simple event reconstruction methods. Furthermore, we 
have explored different schemes to improve on the performance of the methods.
We have shown that the direction and core reconstruction are insensitive to 
details of the weighting of the major axis of the individual telescopes.
The energy resolution benefits from correcting for the height of the
shower maximum. For the analysis of point sources, we get angular, 
core and energy resolutions of 0.22$^\circ$ (0.1$^\circ$), 7.5~m (4.1 m), 
and 22\% (15\%), respectively. The first set of numbers applies to all events
that produce a three-telescope trigger. The set of numbers in brackets
applies to events with a shower core within 150~m from the central telescope and 
with a reconstructed energy exceeding 300 GeV.
In general, our performance estimates for the VERITAS experiment agree well with 
the experimentally verified performance of the H.E.S.S.\ experiment \cite{Ahar:05}.
\\[2ex]
We have compared different $\gamma$-hadron separation methods and have shown that the
information from several parameters can be combined based on the likelihood ratio approach.
Compared to a cut in a single $\gamma$-hadron separation parameter, a cut on the 
likelihood ratio improves the $Q$-factor from 1.7 to 2.6 (three-telescope trigger)
and from 2.4 to 3.6 (four-telescope trigger).

\begin{figure}[bh]
\begin{center}
\includegraphics[width=5.7cm,angle=-90]{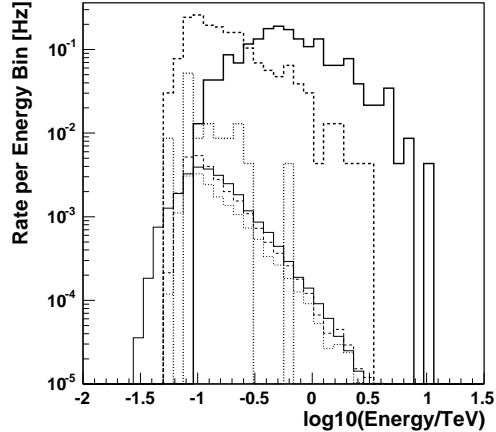}
\end{center}
\caption{\label{fluxes} \normalsize 
The top three curves show the rate of proton events found in a circular
bin of 0.25$^\circ$ radius at the center of the field of view of the telescopes.
The top three curves show the proton rate (i) before application of the
$\gamma$-hadron separation cut as function of the true proton energy
(solid line), (ii) before cut as function of the reconstructed energy
(dashed line), and (iii) after cut as function of the reconstructed energy
(dotted line). 
The lower three lines show the same for the electrons rather than protons.
One can see that the proton background dominates over the electron
background, even after applying the $\gamma$-hadron separation cut. 
We used here the cut 
$\lambda_1(w,\chi_{\rm dir}^{2},\chi_{\rm core}^{2},\chi_{\rm E}^{2})$.
}
\end{figure}

It is instructive to compare the relative importance of the hadron and electron
backgrounds before and after applying the $\gamma$-hadron separation cuts.
Above 30 GeV the BESS energy spectrum for protons impinging on the atmosphere is \cite{Sanu:00}:
\begin{equation}
\frac{dF_{\rm p}}{dE}\,\,=\,\,9.6\times 10^{-9} (E / 1000\,{\rm GeV})^{-2.7} \rm \,cm^{-2}\,s^{-1}\,sr^{-1}\,GeV^{-1}
\end{equation}
The 1 GeV-100 GeV electron spectrum measured by HEAT is \cite{Vern:01}:
\[
\frac{dF_{\rm e}}{dE}\,\,=\,\,1.2\times 10^{-3}\times (E/1\rm\,GeV)^{-1} \times 
\]
\begin{equation}
\hspace*{2cm}
\left(1+(E/5\, \rm GeV)^{2.3}\right)^{-1}\rm \,cm^{-2}\,s^{-1}\,sr^{-1}\,GeV^{-1}
\end{equation}
In Fig.\ \ref{fluxes} we show the proton and electron trigger rates as function of the true and
reconstructed shower energy before and after applying a $\gamma$-hadron separation cut. We used here
the three-telescope trigger condition and a $\gamma$-hadron separation cut 
in $\lambda_1(w,\chi_{\rm dir}^{2},\chi_{\rm core}^{2},\chi_{\rm E}^{2})$.
As proton-initiated showers produce less Cherenkov light than purely electromagnetic showers, 
the reconstructed proton energies are on average lower than the true ones. 
For electrons, there is not such a systematic shift. 
Notwithstanding this dramatic effect, the hadron background still dominates over the 
electron-background -- even after applying the gamma-hadron separation cut.
The result shows that excellent $\gamma$-hadron suppression continues to be of 
utmost importance for VERITAS and other similar experiments. 
\\[2ex]
\hspace*{2cm}\\[2ex]
{\it Acknowledgements:}
HK thanks Jim Buckley, Ira Jung, Scott Hughes, Jeremy Perkins,  
Paul Rebillot and the anonymous referee for fruitful suggestions to improve the text. 
HK acknowledges support of the DOE through the Outstanding Junior Investigator program.
GM acknowledges the support as a Feodor Lynen Fellow of the Alexander 
von Humboldt foundation.
{}

\begin{table}[t]
\begin{center}
\caption{$Q$-factors, photons and proton acceptance and cut value for the $\gamma$-hadron separation methods described in the text. \label{sum}}
\begin{tabular}{|c|c|c|c|c|c|c|c|c|} \hline
$\gamma$-hadron sep. parameter & 
{$Q^a$} & 
{$\epsilon_\gamma$ $\left[\%\right]^a$} & 
{$\epsilon_{\rm p}$ $\left[\%\right]^a$} & 
{cut$^a$} & 
{$Q^b$} & 
{$\epsilon_\gamma$ $\left[\%\right]^b$} & 
{$\epsilon_{\rm p}$ $\left[\%\right]^b$} & 
{cut$^b$}\\ \hline
w                          & 1.51 & 89 & 35  &  0.3  & 2.38 & 46  &  3.7   & -0.1 \\
log($\chi_{\rm dir}^{2})$  & 1.34 & 79 & 35  &  2.7  & 1.53 & 56  &  13    &  2.3\\
log($\chi_{\rm core}^{2})$ & 1.69 & 85 & 25  &  0.9  & 2.30 & 76  &  11    & 0.8\\
log($\chi_{\rm E}^{2}$)    & 1.30 & 86 & 44  & -0.1  & 1.60 & 57  &  13    & -0.5 \\ 
log($\chi_{\rm time}^{2})$ & 1.09 & 85 & 61  &  1.1  & 1.22 & 83  &  47    & 1.1 \\ \hline
$\lambda_1(w,\chi_{\rm core}^{2})$                
                           & 2.27 & 78 & 12 & 0.625 & 3.59 & 62  & 2.9  & 1.375 \\
$\lambda_2(w,\chi_{\rm core}^{2})$                
                           & 2.31 & 76 & 11  & 0.125 & 3.27 & 79  & 6  & 0.125 \\
$\lambda_1(w,\chi_{\rm core}^{2},\chi_{\rm E}^{2})$                
                           & 2.22 & 89 & 18  & 0.375 & 3.09 & 83 & 8.8  & 1.125 \\
$\lambda_2(w,\chi_{\rm core}^{2},\chi_{\rm E}^{2})$                
                           & 2.24 & 63 & 8 & 0.125 & 3.99 & 68 & 2.9  & 0.125 \\
$\lambda_1(w,\chi_{\rm dir}^{2},\chi_{\rm core}^{2},\chi_{\rm E}^{2})$                
                           & 2.60 & 70 & 7.3 & 1.125 & 3.07 & 82 & 7 & 1.375 \\
$\lambda_2(w,\chi_{\rm dir}^{2},\chi_{\rm core}^{2},\chi_{\rm E}^{2})$                
                           & 2.59 & 53 & 4.1 & 0.125 & 3.72 & 81 & 4.7 & -0.125 \\
$\lambda_1(w,\chi_{\rm dir}^{2},\chi_{\rm core}^{2},\chi_{\rm E}^{2},\chi_{\rm time}^{2})$                
                           & 2.40 & 68 &  8 & 1.125  & 3.16 & 64 & 4 & 2.125 \\
$\lambda_2(w,\chi_{\rm dir}^{2},\chi_{\rm core}^{2},\chi_{\rm E}^{2},\chi_{\rm time}^{2})$                
                           & 2.41 & 65 &  7 & -0.125 & 3.17 & 59 & 3.5 & 0.125 \\ \hline
\end{tabular}
\hspace*{0.5cm}$^a${Trigger condition: three or four triggered telescopes.}\\
\hspace*{0.5cm}$^b${Trigger condition: four triggered telescopes.}
\end{center}
\end{table}

%


\end{document}